\documentclass[
]{ceurart}

\pdfoutput=1

\usepackage{mathtools}
\usepackage{graphicx}
\usepackage{booktabs}
\usepackage{svg}
\usepackage{subfig}
\usepackage{multirow}
\usepackage[true]{anonymous}
\usepackage{algorithm}
\usepackage{algorithmic}
\usepackage{amsmath}
\usepackage{breqn}
\usepackage{xcolor}
\usepackage{float}

\usepackage{newfloat}
\usepackage{listings}
\usepackage{listings}
\begin{document}

\copyrightyear{2022}
\copyrightclause{Copyright for this paper by its authors.
  Use permitted under Creative Commons License Attribution 4.0
  International (CC BY 4.0).}

\conference{CIKM MMSR'24: 1st Workshop on Multimodal Search and Recommendations at 33rd ACM International Conference on
Information and Knowledge Management, October 25, 2024, Boise, Idaho, USA}

\renewcommand{\shortauthors}{author name and author name, et al.}

\newcommand{\modelname}{{CAMUE}}

\title{Do We Trust What They Say or What They Do? A Multimodal User Embedding Provides Personalized Explanations}

\author[1]{Zhicheng Ren}[%
email=franklinnwren@g.ucla.edu,
]
\address[1]{Aurora Innovation, 280 N Bernardo Ave, Mountain View, CA 94043
    }
    \cormark[1]

\author[2]{Zhiping Xiao}[%
email=patxiao@uw.edu,
]

\address[2]{University of Washington,
    1410 NE Campus Pkwy, Seattle, WA 98195
    }
\cormark[1]

\author[3]{Yizhou Sun}[%
email=yzsun@cs.ucla.edu,
]

\address[3]{University of California, Los Angeles, Los Angeles, CA 90095
    }

\cortext[1] {All work done at University of California, Los Angeles}
\begin{abstract}
  With the rapid development of social media, the importance of analyzing social network user data has also been put on the agenda.
  User representation learning in social media is a critical area of research, based on which we can conduct personalized content delivery, or detect malicious actors. 
  Being more complicated than many other types of data, social network user data has inherent multimodal nature.
  Various multimodal approaches have been proposed to harness both text (i.e. post content) and relation (i.e. inter-user interaction) information to learn user embeddings of higher quality. The advent of Graph Neural Network models enables more end-to-end integration of user text embeddings and user interaction graphs in social networks. 
  However, most of those approaches do not adequately elucidate which aspects of the data -- text or graph structure information -- are more helpful for predicting each specific user under a particular task, putting some burden on personalized downstream analysis and untrustworthy information filtering.  
  We propose a simple yet effective framework called \textbf{C}ontribution-\textbf{A}ware \textbf{M}ultimodal \textbf{U}ser \textbf{E}mbedding (\textbf{\modelname}) for social networks. 
  We have demonstrated with empirical evidence, that
  our approach can provide personalized explainable predictions, automatically mitigating the impact of unreliable information. 
  We also conducted case studies to show how reasonable our results are. We observe that for most users, graph structure information is more trustworthy than text information, but there are some reasonable cases where text helps more. Our work paves the way for more explainable, reliable, and effective social media user embedding which allows for better personalized content delivery.
\end{abstract}


\maketitle

\section{Introduction}

The advancement of social networks has placed the analysis and study of social network data at the forefront of priorities. User-representation learning is a powerful tool to solve many critical problems in social media studies. Reasonable user representations in vector space could help build a recommendation system~\cite{7355341,10.1145/3308558.3313488}, conduct social analysis \cite{preotiuc-pietro-etal-2017-beyond,Islam_Goldwasser_2022,Jiang_Ren_Ferrara_2023}, detect bot accounts \cite{Varol_Ferrara_Davis_Menczer_Flammini_2017,KUDUGUNTA2018312,Ng_Carley_2023}, and so on. To obtain user-embeddings of higher quality, many multimodal methods are proposed to fully utilize all types of available information from the social networks, including interactive graphs, user profiles, images, and texts from their posts \cite{jin2021heterogeneous,guo2021social,DBLP:journals/corr/abs-1811-02815,huang2019network}. Compared with models using single modality data, multimodal methods utilize more information from the social-media platforms, and hence usually achieve better results in downstream tasks.

Among all modalities in social networks, user-interactive graphs (i.e., what they do) and text content (i.e., what they say) are the two most frequently-used ones, due to their good availability across different datasets and large amount of observations. The graph-neural-network (GNN) models~\cite{kipf2017semisupervised,hamilton2017inductive,velickovic2017graph} makes it more convenient to fuse both the text information and graph-structure information of social-network users, where text-embeddings from language-models such as GloVe \cite{pennington2014glove} or BERT \cite{devlin-etal-2019-bert} are usually directly incorporated into GNNs as node attributes. Although those approaches have achieved great performance in a bunch of downstream tasks \cite{zhou2020graph}, the text information and graph-structure information are fully-entangled with each other, which makes it hard to illustrate the two modalities' respective contributions to learning each user's representation.

It is already found by researchers that different groups of users can behave very differently on social media \cite{6113101}. If such differences are not correctly captured, it might cause significant bias in the user attribute prediction (e.g., political stance prediction) \cite{blank2017digital}. Hence, when learning multi-modal user-representation, it is not only important to ask what the prediction results are, but also important to ask why we are making such predictions for different users (e.g. Are those predictions due to the same reason?). 
Only in that way, we could provide more insights into the user modelings, and potentially enable unbiased and personalized downstream analysis for different user groups.

\begin{figure}[!t]%
    \centering
    \subfloat{{\includegraphics[width=12cm]{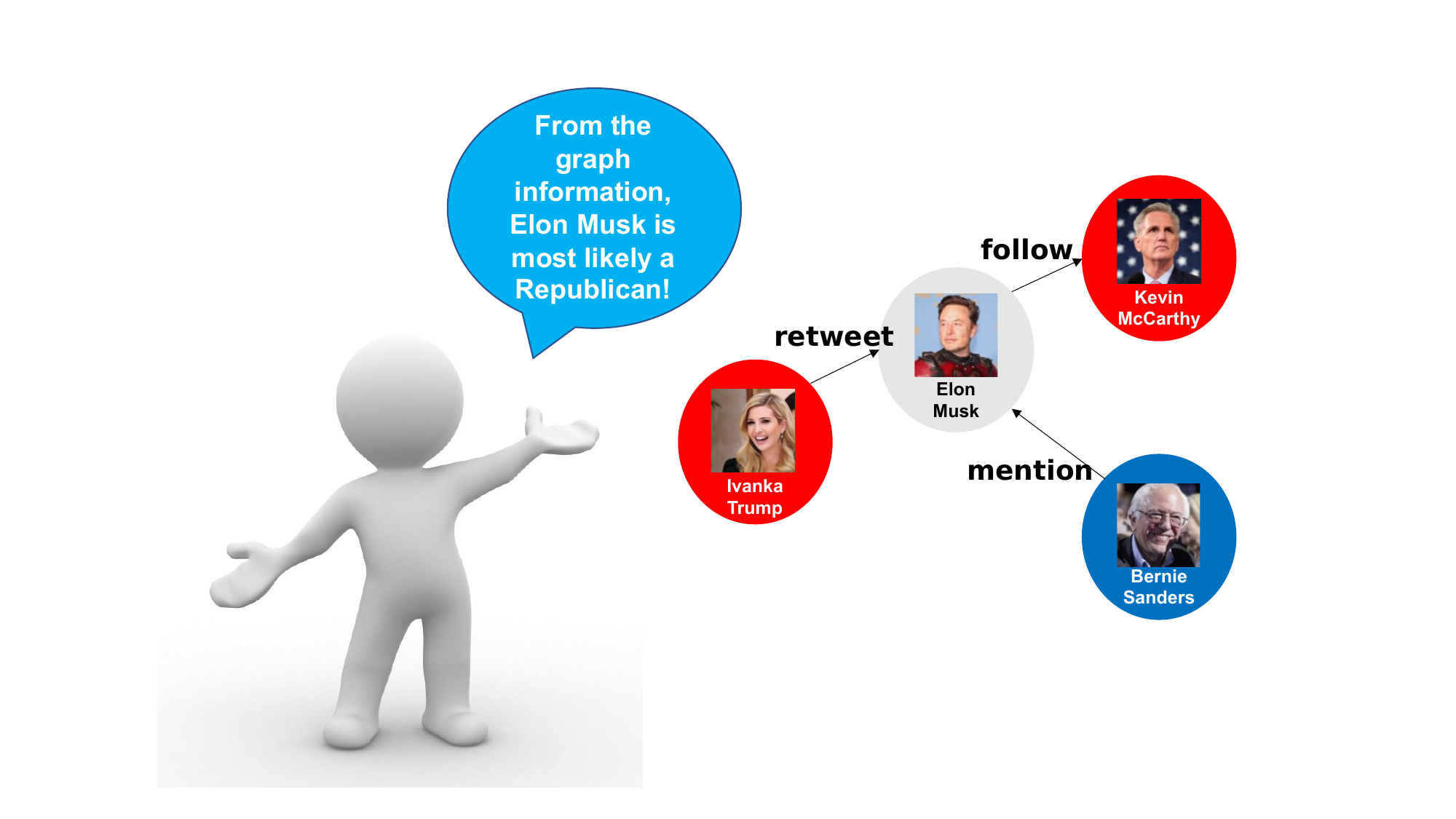} }}%
    \qquad
    \subfloat{{\includegraphics[width=12cm]{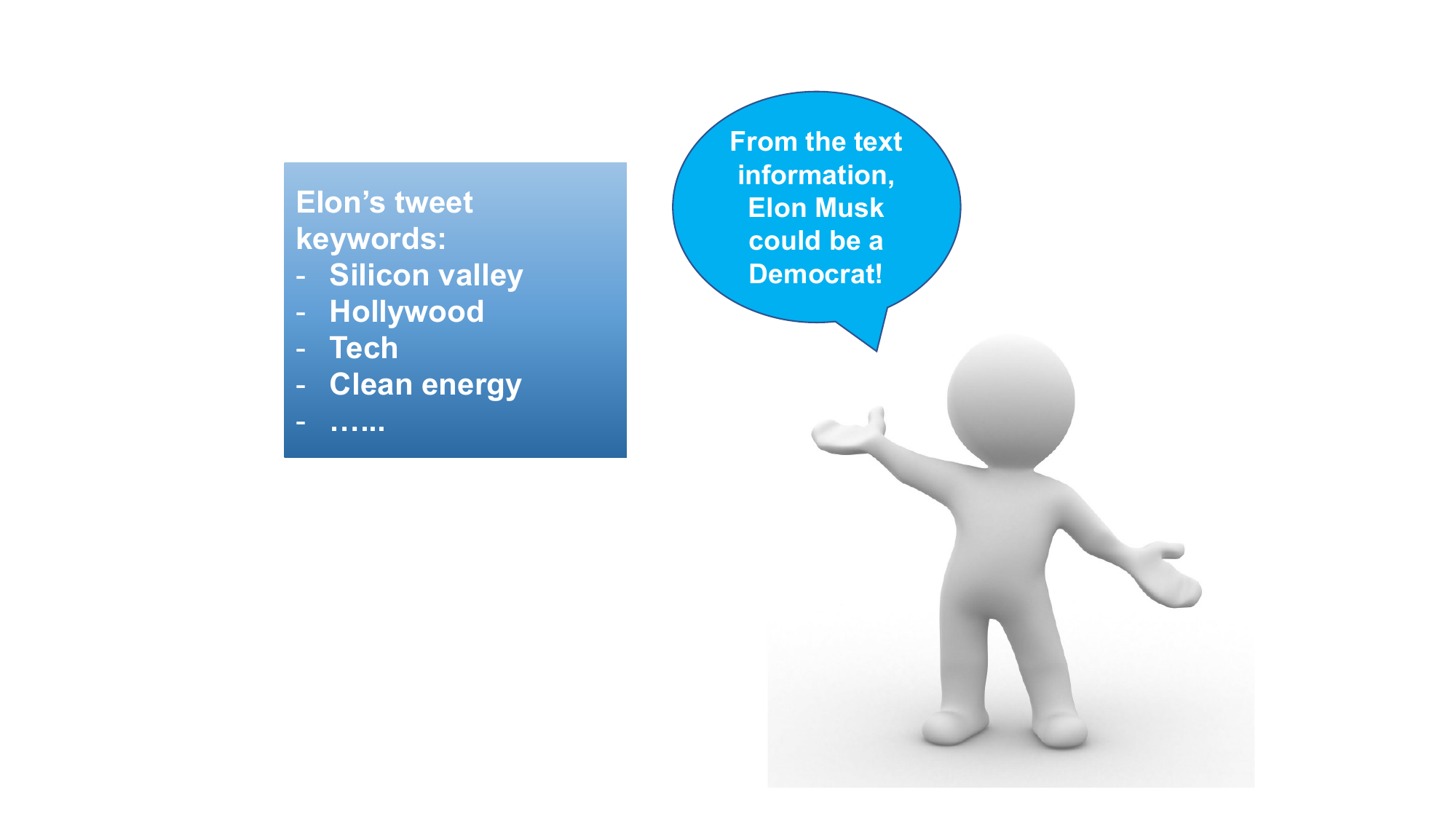} }}%
    \caption{When predicting which political party Elon Musk voted in 2020, graph structure-based methods and text-based methods might reach opposite conclusions. Top: A small subset of Elon's activity with other Twitter users. Bottom: Some keywords extracted from Elon's tweets. All of which are extracted before year 2020.}%
    \label{fig:musk}%
\end{figure}

On the other hand, under a multi-modality setting, if one aspect of a user's data is not trustworthy and misleading, it might still be fused into the model and make the performance lower than single-modality models \cite{10.1145/3394486.3403275}. 
Consider the case when we want to make a political ideology prediction for Elon Musk based on his Twitter content before 2020 U.S presidential election (Figure \ref{fig:musk}), when he has not revealed his clear Republican political-stance yet. If we trust the follower-followee graph structure information, we can see that he is likely to be a Republican since he follows more Republicans than Democrats, and has more frequent interactions with the verified Republicans accounts. However, in his tweet content, his word choice also shows some Democratic traits. Due to the existence of such conflicting information, being able to automatically identify which modality is more trustworthy for each individual becomes essential in building an accurate social media user embedding for different groups of users.

To address the above two shortcomings of text-graph fusion in social networks, we propose a simple yet effective framework called \textbf{C}ontribution-\textbf{A}ware \textbf{M}ultimodal \textbf{U}ser \textbf{E}mbedding (\modelname), which can identify and remove misleading modality from specific social network users during text-graph fusion, in an explainable way. CAMUE uses a learnable attention module to decide whether we should trust the text information or the graph structure information when predicting individual user attributes, such as political stance. Then, the framework outputs a clear contribution map for each modality on each user, allowing personalized explanations for downstream analysis and recommendations. For ambiguous users whose text and graph structure information disagree, our framework could successfully mitigate the unreliable information among different modalities by automatically adjusting the weight of different information accordingly.

We conduct experiments on the TIMME dataset~\cite{10.1145/3394486.3403275} used for a Twitter political ideology prediction task. We observed that our contribution map can give us some interesting new insights. A quantitative analysis of different Twitter user sub-groups shows that link information (i.e., interaction graph) contributes more than text information for most users. This provides insights that political advertising agencies should gather more interaction graph information of Twitter users in the future when creating personalized advertisement content, instead of relying too much on their text data. We also observe that when the graph and text backbone are set to R-GCN and GloVe respectively, our approach successfully ignores the unreliable GloVe embedding and achieves better prediction results. When the text modality is switched to a more accurate BERT embedding, our framework can assign graph/text weights for different users accordingly and achieve comparable performance to existing R-GCN-based fusion methods. We pick $9$ celebrities among the $50$ most-followed Twitter accounts~\footnote{\url{https://socialblade.com/twitter/top/100}}, such as Elon Musk. 
A detailed qualitative analysis of their specific Twitter behaviors shows that our contribution map models their online behaviors well. Finally, we run experiments on the TwiBot-20-Sub dataset~\cite{Feng2021TwiBot20AC} used for a Twitter human/bot classification task, showing that our framework could be generalized to other user attribute prediction tasks. By creating social media user embeddings that are more explainable, reliable, and effective, our framework enables improved customized content delivery.


\section{Preliminaries and Related Work}

\subsection{Multimodal Social Network User Embedding} 

Social network user embedding is a popular research field that aims to build accurate user representations. A desirable user embedding model should accurately map sparse user-related features in high-dimensional spaces to dense representations in low-dimensional spaces. Multimodal social network user embedding models utilize user different types of user data to boost their performance. Commonly-seen modality combinations include graph-structure (i.e. link) data and text data \cite{ribeiro2018characterizing}, graph-structure data and tabular data \cite{zhang2018anrl,8326519,jin2021heterogeneous}, and graph-structure data, text data and image data altogether~\cite{zhang2018user} \cite{ni2022two}, etc.

Among those multi-modality methods, the fusion of graph-structure data and text data has always been one of the mainstream approaches for user embedding. At an earlier stage, without much help from the GNN models, most works trained the network-embedding and text-embedding separately and fused them using a joint loss \cite{ijcai2019p881,DBLP:journals/corr/abs-1805-04612,10.1145/2832907,benton2016learning}. With help of the GNN models, a new type of fusion method gained popularity, where the users' text-embeddings are directly incorporated into GNNs as node attributes~\cite{ribeiro2018characterizing, yao2019graph, DBLP:journals/corr/abs-2104-12259}.

Despite their good performance, all existing models did not explain how much the graph structure or text information of a particular user contributed to its final prediction result, make it difficult to give customized modality weight for downstream analysis or recommendations. Also, if one modality is very poorly learned, it can be counter-effective to the user embedding quality, make it even worse than their single-modality counterparts \cite{10.1145/3394486.3403275}. How to address this problem in a universally-learned way instead of heuristic-based information filtering, has largely gone under-explored. Hence, we propose a framework that would not only utilize both text and graph-structure information together but also reveal their relative importance along with our prediction result.

\subsection{Graph Neural Network} 

Graph Neural Network (GNN) refers to a collection of deep learning models that learn node embedding through iterative aggregation of information from neighboring nodes, using a convolutional operator. The majority of the GNN architectures include a graph convolution layer in a form that can be characterized as message-passing and aggregation. A general formula for such convolution layers is:

\begin{equation}
    \textbf{H}^{(l)} = \sigma(\Tilde{\textbf{A}}\textbf{H}^{(l-1)}\textbf{W}^{(l)})\,,
\end{equation}
where $\textbf{H}^{(l)}$ represents the hidden node representation of all nodes at layer $l$, operator $\sigma$ is a non-linear activation function, and the graph-convolutional filter $\Tilde{\textbf{A}}$ is a matrix that usually takes the form of a transformed (e.g., normalized) adjacency matrix $\textbf{A}$, and the layer-$l$'s weight $\textbf{W}^{(l)}$ is learnable.

In the past few years, GNN models have reached SOTA performances in various graph-related tasks, and are widely regarded as a very promising technique to generate node embedding for users in social-network graphs.~\cite{kipf2017semisupervised,schlichtkrull2018modeling,hamilton2017inductive,velickovic2017graph}.



\subsection{Neural Network-based Language Models} 

The field of natural language processing has undergone a significant transformation with the advent of neural-network-based language models. Word2Vec \cite{NIPS2013_9aa42b31} introduced two architectures: Continuous Bag-of-Words (CBOW) and Skip-Gram. CBOW predicts a target word given its context, while Skip-Gram predicts context words given a target word. GloVe \cite{pennington2014glove} model went beyond by incorporating global corpus statistics into the learning process. ELMo \cite{peters-etal-2018-deep} was another significant step forward, as it introduced context-dependent word representations, making it possible for the same word to have different embeddings if the context is different. BERT \cite{devlin-etal-2019-bert} is a highly influential model that is built on the transformer architecture~\cite{vaswani2017attention}, pre-trained on large text corpora using, for example, masked language modeling and next-sentence prediction tasks. Recently, large language models like GPT-3 \cite{NEURIPS2020_1457c0d6}, InstructGPT \cite{10.5555/3600270.3602281}, and ChatGPT have achieved significant breakthroughs in natural-language-generation tasks. All of those large language models (LLMs) are frequently used to generate text embedding for social network users. 

Our framework does not rely on any specific language model, and we do not have to use LLMs. Instead, we use language-models as a replaceable component, making it possible for either simpler ones like GloVe or more complicated ones like BERT to fit in. We will explore some different options in the experimental section.

\subsection{Multimodal Explanation Methods}
In the past, several methods have been proposed to improve the interpretability and explainability of multimodal fusions \cite{9391727}. Commonly used strategies include attention-based methods \cite{DBLP:journals/corr/abs-2011-13681,DBLP:journals/corr/GoyalMPB16,10.1007/978-3-030-33850-3_3}, counterfactual-based methods \cite{10.1145/3442188.3445899, DBLP:journals/corr/abs-1806-09809}, scene graph-based methods \cite{alipour2020impact} and knowledge graph-based methods \cite{9357868}. Unfortunately, most of them focus on the fusion of image modality and text modality, primarily the VQA task, while to the best of our knowledge, no work focuses on improving the explainability between the network structure data and text data in social-network user embedding.

\section{Problem Definition}
Our general goal is to propose a social network user embedding fusion framework that could answer: 1. which modality (i.e. text or graph structure, saying or doing) contributes more to our user attribute prediction, hence allowing more customized downstream user behavior analysis and 2. which modality should be given more trust for each user, and automatically filter out the untrustworthy information when necessary, in order to achieve higher-quality multi-modal user-embedding.

\subsection{Problem Formulation} A general framework of our problem could be formulated as follows: given a social media interaction graph $\mathcal{G} = (\mathcal{V},\mathcal{E})$ with node set $\mathcal{V}$ representing users and edge set $\mathcal{E}$ representing links between users. Let $\textbf{X} = [x_1, x_2, x_3, \cdots, x_n]$ be the text content of $n = |\mathcal{V}|$ users, $\textbf{Y} = [y_1, y_2, y_3, \cdots, y_n]$ be the labels of those users, $\textbf{A} = [ \textbf{A}^1, \textbf{A}^2, \cdots, \textbf{A}^m]$ be the adjacency matrices of $\mathcal{G}$, $m$ be the number of link types and $\textbf{A}^i \in \mathbb{R}^{n \times n}$, our training objective is:

\begin{equation}
\text{min} \ \mathbb{E} \left[ \mathcal{L}
\left(f \left(\mathcal{G}, \textbf{X}\right), \textbf{Y}\right) \right]
\end{equation}

Here, $\mathcal{L}$ is the loss of our specific downstream task, and $f$ is some function that combines the graph structure information and text information, producing a joint user embedding.

\subsection{Preliminary Experiment} To investigate the effectiveness of the existing GNN-based multimodal fusion methods in filtering the unreliable modality when the graph structure and text contradict, we run experiments using a common fusion method that feeds the fine-tuned BERT features into the R-GCN backbone, similar to the approaches in \cite{ribeiro2018characterizing} and \cite{guo2021social}. We observe that this conventional fusion method fails to filter the unreliable information for some of those ambiguous users. Table \ref{tab:table1} show two politicians whose Twitter data contains misleading information, either in the graph structure or text data. While the single-modality backbones which are trained without misleading information give the correct predictions, the multi-modality fusion method is fooled by the misleading information and is not able to make correct predictions.

\begin{table}[ht]
\centering
\caption{Left: A snippet for former U.S. Rep. Ryan Costello, his tweet text content is informative for the political Party prediction, but his Twitter interaction graph data could be misleading. \\
Right: A snippet for U.S. Senator Sheldon Whitehouse, his Twitter interaction graph data is informative for the political Party prediction, but his tweet text content could be misleading.}

\begin{minipage}{0.5\textwidth}
        \centering
    \begin{tabular}{|p{2.4cm}|p{3.8cm}|}
        \hline
        \textbf{Name} & Ryan Costello \\
        \hline
        \textbf{Ground Truth Party} & Republican \\
        \hline
        \textbf{Sample Graph Data} & Liked Ben Rhodes (Democrat) 20 times.
        
        Liked Donald Trump 0 time.
        
        Following Mike Quigley (Democrat). \\
        \hline
        \textbf{Sample Text Data} & Despite Trump, Iran's elections \& chaotic ME, some Democrats want to race ahead with ill-conceived Iran sanctions
        
        RT @SaeedKD: Iran's people care about elections. The so-called democratic fringe doesn't - by me\\
        \hline
        \textbf{Graph-backbone Prediction} & Democrat (Wrong) \\
        \hline
        \textbf{Text-backbone Prediction} & Republican (Right) \\
        \hline
        \textbf{Fused Model Prediction} & Democrat (Wrong) \\
        \hline
    \end{tabular}
\end{minipage}%
\hfill
\begin{minipage}{0.5\textwidth}
        \centering
    \begin{tabular}{|p{2.4cm}|p{3.8cm}|}
        \hline
        \textbf{Name} & Sheldon Whitehouse \\
        \hline
        \textbf{Ground Truth Party} & Democrat \\
        \hline
        \textbf{Sample Graph Data} & Liked Senate Democrats Official Account 26 times.
        
        Not following Donald Trump.
        
        Following Barack Obama. \\
        \hline
        \textbf{Sample Text Data} & My Republican partner on the CARA bill, @SenRobPortman, writes a powerful editorial on the success of CARA and CURES (which provided a needed boost of funding to match CARA).
        
        Good move by Trump Administration. Cong. @JimLangevin \& \\
        \hline
        \textbf{Graph-backbone Prediction} & Democrat (Right) \\
        \hline
        \textbf{Text-backbone Prediction} & Republican (Wrong) \\
        \hline
        \textbf{Fused Model Prediction} & Republican (Wrong) \\
        \hline
    \end{tabular}
\end{minipage}
\label{tab:table1}
\end{table}

These insights revealed the importance of having a more flexible and explainable framework for learning multimodal user embedding.

\section{Methodology}

We propose a framework of \textbf{C}ontribution-\textbf{A}ware
\textbf{M}ultimodal \textbf{U}ser \textbf{E}mbedding (\modelname), a fusion method for text data and graph structure data when learning user embedding in social networks. The key ingredient of this framework is an attention gate-based selection module which is learned together with the link and text data, and decide which information we want to trust more for each particular user.

\begin{figure}%
    \centering
    \includegraphics[width=0.7\linewidth]{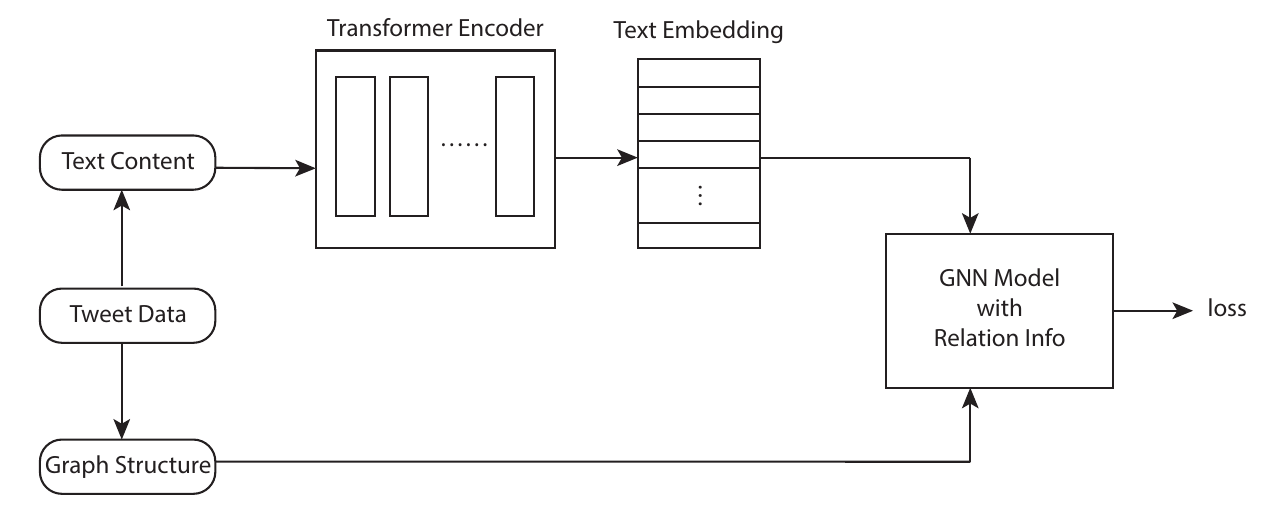}
    \includegraphics[width=0.7\linewidth]{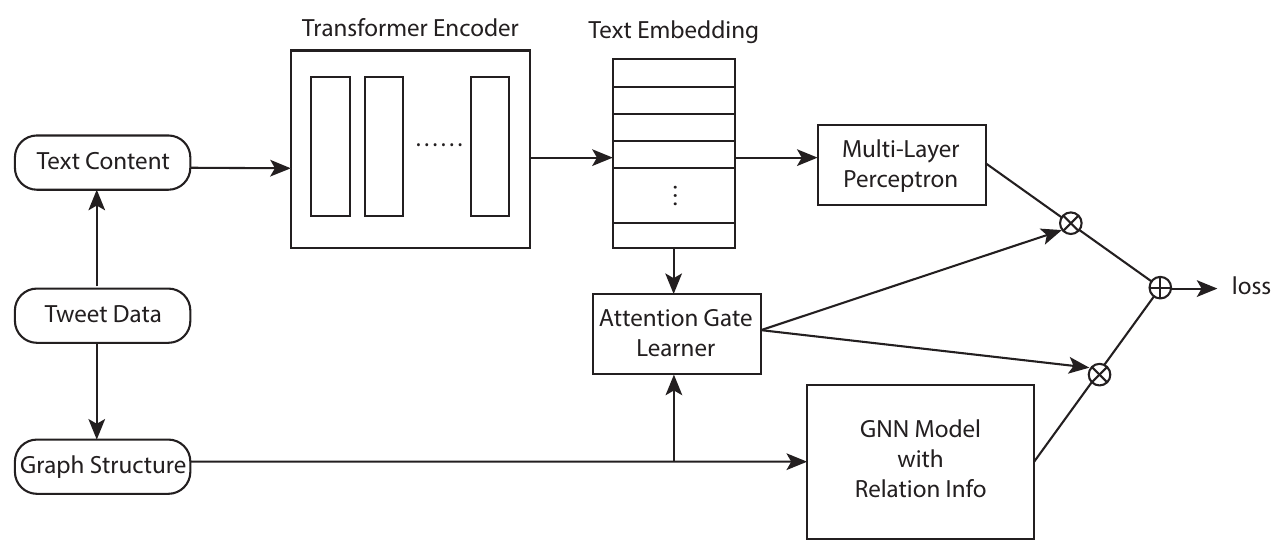}
    \caption{The architectures of our framework. Top: Simple Fusion method for GNN (baseline), bottom: {\modelname}}%
    \label{fig:method}%
\end{figure}

Our framework has three main parts: a text encoder, a graph encoder, and an attention-gate learner. The text content of each user passes through the text encoder and generates a text embedding for that user. The embedding is then passed through a three-layer MLP for fine-tuning. The adjacency matrix of the users passes through the graph encoder and generates a node embedding for that user. At the same time, both the text embedding and the graph adjacency matrix pass through our attention gate learner. The output of this module is two attention weights, $\alpha$ and $\beta$, which control the proportion of our graph structure information and text information. Without loss of generality, if we make R-GCN our graph encoder and BERT our text encoder, our model will be trained in the following way (Equation 3-6, also illustrated in Figure \ref{fig:method}):

\begin{equation}
    \textbf{H}^{(1)} = \sigma(\text{concat}(\textbf{A}^1 + \textbf{A}^2 + \cdots + \textbf{A}^m, \text{BERT}emb\left(\textbf{X}\right)\textbf{W}^{(1)})
\end{equation}

\begin{equation}
    \textbf{H}^{(2)} = \sigma(\textbf{H}^{(1)} \textbf{W}^{(2)})
\end{equation}

\begin{equation}
    \left[{e}_\alpha, {e}_\beta\right] = \textbf{H}^{(2)} \textbf{W}^{(3)}
\end{equation}

\begin{equation}
    \alpha = \text{softmax}({e}_\alpha), \ \beta = \text{softmax}({e}_\beta)
\end{equation}

Where $\textbf{H}$ and $\textbf{W}$ are hidden layers and weights of our attention gate learner, $\textbf{X} = [x_1, x_2, x_3, \cdots, x_n]$ is the text content, $\text{BERT}emb$ is the BERT encoding module, $\textbf{A} = [ \textbf{A}^1, \textbf{A}^2, \cdots, \textbf{A}^m]$ is the adjacency matrices of $\mathcal{G}$ and $m$ is the number of link types.

Then, our overall training objective becomes:

\begin{align*}
    \text{min} \ \mathbb{E} [ &\mathcal{L} ((\alpha + \lambda)\ \text{R-GCN}emb(\mathcal{G}) \\
    &+ (\beta +\lambda)\ \text{BERT}emb(\textbf{X}), \textbf{Y}
    ) ]
\end{align*}

Here, $\lambda$ acts as a regularizer to ensure our model is not overly dependent on a single modality.

Our methods offer two levels of separation. First, we separate the text encoder and graph encoder to allow better disentanglement on which data contributes more to our final prediction results. Second, we separate the learning of the downstream tasks and the learning of which data modality (i.e. text or graph structure) we can rely on more. This makes our framework adaptable to different downstream social media user prediction tasks. The learned trustworthiness of different modalities allows for auto-adjustment of the weight between graph structure and text modalities, hence filtering any unreliable information once they are discovered.

Figure \ref{fig:method} shows the overall architecture of our framework, note that the graph structure encoder and text encoder could be replaced by any other models which serve the same purposes.

We give a short complexity analysis of our architecture for the case of R-GCN + BERT: Since we are using sparse adjacency matrix for R-GCN, the graph encoder part has a complexity of $\mathcal{O}(L_{\text{graph}}EF_{\text{graph}} + L_{\text{graph}}NF_{\text{graph}}^2)$ (according to \cite{blakely2021time}), where $L$ is the number of layers, $E$ is the number of edges, $N$ is the number of nodes, and $F$ is the feature dimension. Since we fixed the maximum text length to be a constant for the text encoder, it has a complexity of $\mathcal{O}(F_{\text{text}}^2)$ (based on \cite{vaswani2018tensor2tensor}). Since $F_{\text{text}}$ and $F_{\text{graph}}$ are about comparable size, our fusion module has the complexity of $\mathcal{O}(F_{\text{graph}}^2 + F_{\text{text}}^2)$, so the overall complexity is $\mathcal{O}(L_{\text{graph}}EF_{\text{graph}} + L_{\text{graph}}NF_{\text{graph}}^2 + F_{\text{text}}^2)$, hence we are not adding extra time complexity.

\section{Experiments}

\subsection{Tasks and Datasets}
We run experiments on two Twitter user prediction tasks: 1. Predicting the political ideology of Twitter users (Democrat vs Republican) and 2. Predicting whether a Twitter user account is a human or a bot.

\subsubsection{TIMME} TIMME \cite{10.1145/3394486.3403275} introduced a multi-modality Twitter user dataset as a benchmark of political ideology prediction task for Twitter users. TIMME contains $21,015$ Twitter users and $6,496,112$ Twitter interaction links. Those links include follows, retweets, replies, mentions, and likes, together they form a large heterogeneous social network graph. TIMME also contains $6,996,310$ raw Twitter content from those users. Hence, it will be a good dataset to study different fusion methods of text features and graph structure features. In TIMME, there are $586$ labeled politicians and $2,976$ randomly sampled users with a known political affiliation. Some of them are ambiguous users we investigated before. Labeled nodes belong to either Democrats or Republicans. Note that the dataset cut-off time is 2020, so the political polarity of many public figures (e.g. Elon Musk) have not been reviewed at that time.
\subsubsection{TwiBot-20-Sub} TwiBot-20 \cite{Feng2021TwiBot20AC} is an extensive benchmark for Twitter bot detection, comprising $229,573$ Twitter accounts, of which $11,826$ are labeled as human users or bots. The dataset also contains $33,716,171$ Twitter interaction links and $33,488,192$ raw Twitter content. The links in TwiBot-20 include follows, retweets, and mentions. To further examine the generalizability of our method, we run experiments for Twitter bot account detection on the TwiBot-20 dataset. To reduce the computation cost of generating node features and text features, we randomly subsample $3,000$ labeled users and $27,000$ unlabeled users from the TwiBot-20 dataset, and form a new dataset called TwiBot-20-Sub. In this way, the size and label sparsity of the TwiBot-20-Sub dataset becomes comparable with the TIMME dataset.

\subsubsection{Train-test Split}
We split the users of both datasets into an $80\%$:$10\%$:$10\%$ ratio for the training set, validation set, and test set respectively.

\subsection{Implementation Detail}

To test the effectiveness of our framework across different models, we choose two single-modality text encoders, GloVe and BERT, and two single-modality graph encoders, MLP and R-GCN.

The GloVe embedding refers to the Wikipedia
2014 + Gigaword 5 (300d) pre-trained version.  \footnote{glove.6B.zip from \url{https://nlp.stanford.edu/projects/glove/}} The BERT embedding refers to the sentence level ($[CLS]$ token) embedding of BERT-base model~\cite{47751} after fine-tuning the pre-trained model's parameters on the tweets from our training set consisting of $80\%$ of the users. We chose a max sequence length of 32. After the encoding, we have a 300-dimension text embedding for GloVe and a 768-dimension text embedding for BERT.

We choose a modified version of R-GCN from TIMME \cite{10.1145/3394486.3403275} as an R-GCN graph encoder. R-GCN \cite{schlichtkrull2018modeling} is a GNN model specifically designed for heterogeneous graphs with multiple relations. In the TIMME paper, it is discovered that assigning different attention weights to the relation heads of the R-GCN model could improve its performance. Hence, we adopt their idea and use the modified version of R-GCN. We did not use the complete TIMME model since it is designed for multiple tasks which is outside our research scope, and will overly complicate our model.

We also choose a 3-layer MLP as another graph encoder for comparison, the adjacency list for each user is passed to the MLP.



Large language models (LLMs) like ChatGPT are powerful in understanding texts, but they usually have a great number of parameters, making traditional supervised fine-tuning a hard and costly task \cite{NEURIPS2020_1457c0d6}. Instead, less resource-intensive methods like few-shot learning, prompt tuning, instruction tuning, and chain-of-thought are more frequently used to adapt LLMs on specific tasks \cite{longpre2023flan}. We do not use large language models as one of the options for the text encoder since those methods are not compatible with our framework -- they do not provide a well-defined gradient to train our attention gate learner.

We run experiments on a single NVIDIA Tesla A100 GPU. We used the same set of hyper-parameters as in the TIMME paper, with the learning rate being 0.01, the number of GCN hidden units being 100, and the dropout rate being 0.1, on a PyTorch platform. For a fair comparison, we run over 10 random seeds for each algorithm on each task.

\section{Results and Analysis}

\subsection{Contribution Map}

To show that our framework could essentially provide personalized explanations during the fuse of modalities, we draw the contribution map based on $\alpha$ (graph weight) and $\beta$ (text weight) attention for users from each dataset. The darker the color is, the weight of the corresponding modality is closer to $1$. In the contribution map, pure white indicates a zero contribution ($0$) from a modality, while pure dark blue indicates a full contribution ($1$).

The top figure of Figure \ref{fig:attention} shows the contribution map output when the text encoder is BERT and the graph encoder is R-GCN, on a subgroup of the TIMME dataset consisting of some politicians and some random Twitter users. To avoid any misuse of personal data information, we hide the names of random Twitter users and only include politicians whose Twitter accounts are publically available at \footnote{\url{https://tweeterid.com/}}. As we can see, there is a clear cut between the percentage of contributions from different modalities to the final prediction. It is notable that for the two ambiguous politician users we mentioned earlier (Ryan Costello and Sheldon Whitehouse), {\modelname} could give correct attention, where we should trust more text data from Mr. Costello while trusting more graph structure data from Mr. Whitehouse.

The bottom figure of Figure \ref{fig:attention} shows the contribution map output when the text encoder is GloVe and the graph encoder is R-GCN, on the same subgroup of the TIMME dataset. Note that for all shown users text information does not contribute to the final prediction. This could be attributed to the fact that GloVe is not very powerful for sentence embedding, especially when the text is long. This contribution map shows us our framework filters out the text modality almost completely when it is not helpful for our user embedding learning. As we can see from table \ref{tab:performances_conclusion}, the traditional fusion method for GloVe+R-GCN only yields an accuracy of $0.840$, which is much lower than the single graph structure modality prediction ($0.953$) using R-GCN, due to unreliable GloVe embedding. In contrast, our CAMUE method obtains a higher accuracy ($0.954$) than the single modality models by disregarding the unreliable information.

\begin{figure*}%
    \centering
    \subfloat{{\includegraphics[width=13cm]{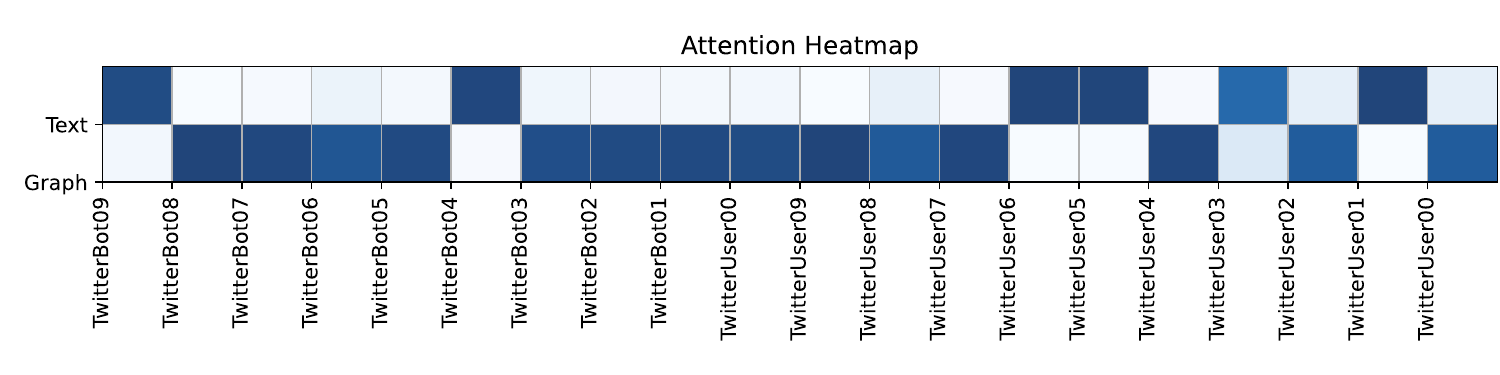} }}%
    \caption{Contribution map for TwiBot-20-Sub dataset, \modelname(BERT, R-GCN)}%
    \label{fig:attention_2}%
\end{figure*}

Figure \ref{fig:attention_2} shows the contribution map output for the same set of encoders on a subgroup of the Twibot-20-Sub dataset. There is also a clear cut between the percentage of contributions from different modalities, for both the human Twitter accounts and bot accounts.

Hence, we verify that our framework could both provide personalized modality contribution and drop low-quality information during the fuse of modalities. Some quantitative analysis of how this low-quality information filtering could benefit the general model performance could be found in the next section, and some qualitative analysis about what new insights we could gain from the output of our framework could be found in the case study section.

\subsection{General Performance}

Table~\ref{tab:performances_conclusion} shows the performance of {\modelname} on different combinations of encoders. The traditional fusion method in Figure \ref{fig:method} is denoted as ``simple fusion''. For MLP, we do not have such a natural fusion method. We also add ``{\modelname}, fixed params'' as an ablation experiment to prove the effectiveness of our attention gate-based selection module.

\begin{figure*}%
    \centering
    \subfloat{{\includegraphics[width=13cm]{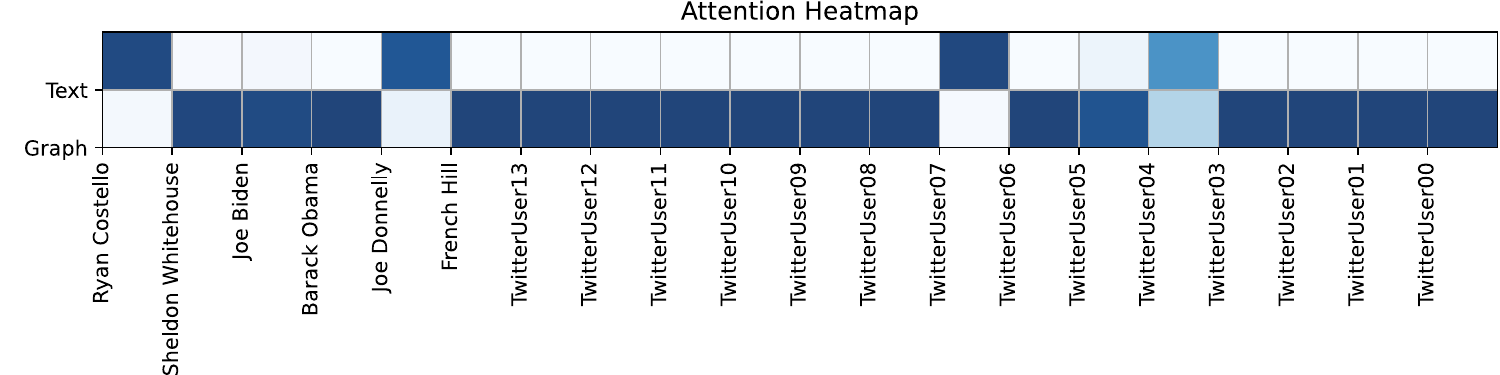} }}%
    \qquad
    \subfloat{{\includegraphics[width=13cm]{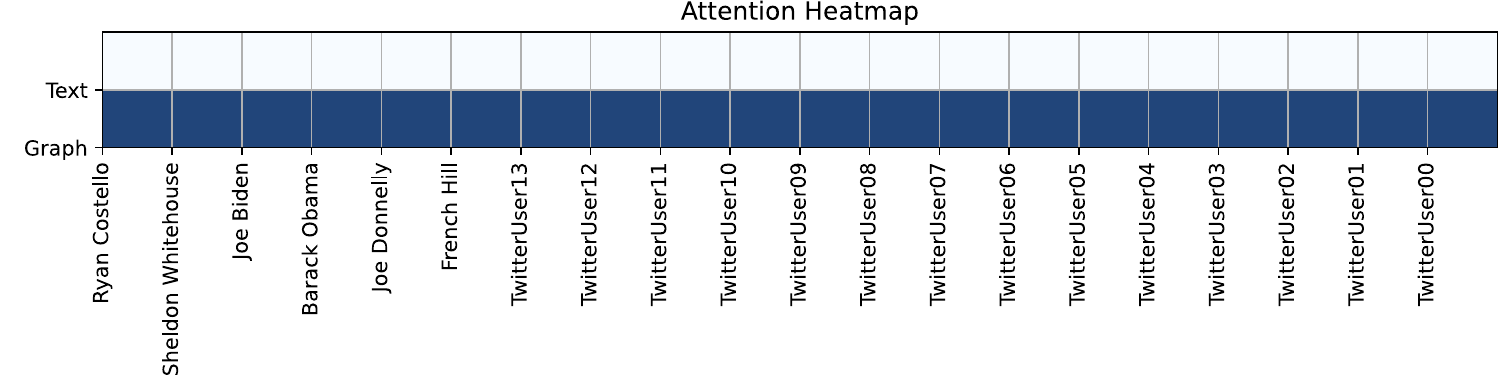} }}%
    \caption{Contribution map for TIMME dataset, top: \modelname(BERT, R-GCN), bottom: \modelname(GloVe, R-GCN)}%
    \label{fig:attention}%
\end{figure*}

We observe that within those combinations, sometimes simple fusion methods are significantly worse than single-modality methods (e.g. GloVe+R-GCN vs R-GCN only) due to some untrustworthiness in one of the modalities. However, any fusion under our {\modelname} framework always performs better than their respective single modality methods. That suggests that our algorithm can benefit from attending to the more reliable modality between text and graph structure, if one particular modality is not trustworthy (e.g. GloVe embedding), and learning not to consider it when making predictions (as we can see in Figure \ref{fig:attention}, bottom).

It is also notable that our {\modelname} method outperforms ``{\modelname}, fixed params''. These results suggest that adjusting the weight of different modalities dynamically yields better performance than fixed weights of modalities. Finally, when the text modality is
switched to a more accurate BERT embedding, our framework still gives comparable performance to its corresponding simple fusion methods.


\begin{table}[h]
\caption{The overall performance (format: accuracy ; f1-score) of {\modelname} on different social media data sets.}
\begin{tabular}{c|ll|cc}
\hline
\multirow{2}{*}{Algorithm}                                                       & \multicolumn{2}{c|}{Encoder Variant}                                           & \multicolumn{2}{c}{Data Set}                       \\ \cline{2-5} 
                                                                                 & \multicolumn{1}{c|}{Text}                     & \multicolumn{1}{c|}{Graph} & \multicolumn{1}{c|}{TIMME}         & TwiBot-20-Sub \\ \hline
\multirow{2}{*}{text-only}                                                       & \multicolumn{1}{l|}{GloVe}                  & \multirow{2}{*}{N/A}     & \multicolumn{1}{c|}{0.688 ; 0.681} & 0.565 ; 0.511 \\ \cline{2-2} \cline{4-5} 
                                                                                 & \multicolumn{1}{l|}{BERT}                   &                          & \multicolumn{1}{c|}{0.862 ; 0.859} & 0.731 ; 0.722 \\ \hline
\multirow{2}{*}{link-only}                                                       & \multicolumn{1}{l|}{\multirow{2}{*}{N/A}}   & MLP                      & \multicolumn{1}{c|}{0.932 ; 0.930} & 0.707 ; 0.697 \\ \cline{3-5} 
                                                                                 & \multicolumn{1}{l|}{}                       & R-GCN                    & \multicolumn{1}{c|}{0.953 ; 0.953} & 0.735 ; 0.728 \\ \hline
\multirow{2}{*}{simple fusion}                                                   & \multicolumn{1}{l|}{GloVe}                  & R-GCN                    & \multicolumn{1}{c|}{0.840 ; 0.837} & 0.683 ; 0.675 \\ \cline{2-5} 
                                                                                 & \multicolumn{1}{l|}{BERT}                   & R-GCN                    & \multicolumn{1}{c|}{0.959 ; 0.959} & 0.791 ; 0.787 \\ \hline
\multirow{4}{*}{\begin{tabular}[c]{@{}c@{}}{\modelname} w. \\ fixed params\end{tabular}} & \multicolumn{1}{l|}{\multirow{2}{*}{GloVe}} & MLP                      & \multicolumn{1}{c|}{0.938 ; 0.937} & 0.700 ; 0.691 \\ \cline{3-5} 
                                                                                 & \multicolumn{1}{l|}{}                       & R-GCN                    & \multicolumn{1}{c|}{0.952 ; 0.951} & 0.734 ; 0.727 \\ \cline{2-5} 
                                                                                 & \multicolumn{1}{l|}{\multirow{2}{*}{BERT}}  & MLP                      & \multicolumn{1}{c|}{0.940 ; 0.938} & 0.732 ; 0.722 \\ \cline{3-5} 
                                                                                 & \multicolumn{1}{l|}{}                       & R-GCN                    & \multicolumn{1}{c|}{0.952 ; 0.951} & 0.779 ; 0.771 \\ \hline
\multirow{4}{*}{{\modelname}}                                                            & \multicolumn{1}{l|}{\multirow{2}{*}{GloVe}} & MLP                      & \multicolumn{1}{c|}{0.945 ; 0.944} & 0.707 ; 0.697 \\ \cline{3-5} 
                                                                                 & \multicolumn{1}{l|}{}                       & R-GCN                    & \multicolumn{1}{c|}{0.954 ; 0.953} & 0.738 ; 0.731 \\ \cline{2-5} 
                                                                                 & \multicolumn{1}{l|}{\multirow{2}{*}{BERT}}  & MLP                      & \multicolumn{1}{c|}{0.935 ; 0.933} & 0.744 ; 0.738 \\ \cline{3-5} 
                                                                                 & \multicolumn{1}{l|}{}                       & R-GCN                    & \multicolumn{1}{c|}{0.961 ; 0.960} & 0.782 ; 0.776 \\ \hline
\end{tabular}
\label{tab:performances_conclusion}
\end{table}

\subsection{Case Studies}

\textbf{User Sub-groups} Table \ref{tab:table12} gives a quantitative analysis when the text encoder is BERT and the graph encoder is R-GCN, for different sub-groups of Twitter users we are interested in. In general, graph structure information contributes the most when it comes to bot accounts. One possible explanation for this is the variety of bot accounts on Twitter, such as those for business advertising, political outreach, and sports marketing \cite{Feng2021TwiBot20AC}. Bots with different usage might talk very differently, however, they may share some common rule-based policies when trying to interact with humans on Twitter \cite{alothali2018detecting, 10.1145/3292522.3326015}.

Graph structure information contributes the second highest when it comes to politicians. This is also not surprising since politicians are generally more inclined to retweet or mention events related to their political parties \cite{10.1145/3394486.3403275}. It is also notable that the weight of text information for Republicans is slightly less than it is for Democrats. This aligns with the findings in \cite{xiao2023detecting} that Democrats have a slightly more politically polarized word choice than Republicans.

For random users, the weight of text information is the largest, although still not as large as the weight of graph structure information. This could be attributed to the pattern that many random users interact frequently with their non-celebrity families and friends on Twitter, who are more likely to be politically neutral.

\begin{table}[!ht]
\caption{$\%$ Users whose $\alpha > \beta$ for different subgroups}
\centering
\begin{tabular}{ c | c }
\hline
Subgroup              & $\%$ Users \\
\hline
Democrats & 70.9   \\
Republicans & 76.1   \\
Politicians & \underline{76.2}   \\
Non-politicians with Party affiliations      & 72.4        \\
Non-bot random users      & 61.2      \\
Bot accounts     & \textbf{77.3}     \\
TIMME, aggregated      & 73.5      \\
TwiBot-20-Sub, aggregated     & 70.1     \\

\hline
\end{tabular}
    \label{tab:table12}
    \end{table}

\begin{table}[h]
\caption{Predicted Political Stance of Some News Agencies}
\centering
\begin{tabular}{ c | c | c }
\hline
News Agency & Prediction & Text or Graph\\
\hline
New York Times & D & Graph\\
Washington Post & D & Graph\\
Wall Street Journal & R & Text\\
USA Today & D & Graph\\
CNN & D & Graph\\
Fox News & R & Text\\
Guardian & D & Text\\
Associated Press & R & Graph\\
US News & D & Graph\\
MSNBC & D & Graph\\
BBC & R & Graph\\
National Review & R & Graph\\
Bloomberg & D & graph\\
\hline
\end{tabular}
\label{tab:table11}
\end{table}

Table \ref{tab:table11} shows some predicted political stances and the main contributing modalities of a group of news agencies. We could see that the majority of them have more reliable graph structure information than text information. This is not surprising since most news agency tends to use neutral words to increase their credibility, hence it is hard to gather strong political stances from their text embedding, except for some of them like Fox News and Guardian which are known to use political polarized terms more often \cite{xiao2023detecting, doi:10.1177/0263395720955036}. Our framework is able to capture this unique behavior pattern for Fox News and Guardian, meanwhile giving mostly accurate political polarity predictions aligning with results in \cite{xiao2023detecting} and \footnote{\url{https://www.allsides.com/media-bias/ratings}}.

\begin{figure*}%
    \centering
    \subfloat{{\includegraphics[width=16cm]{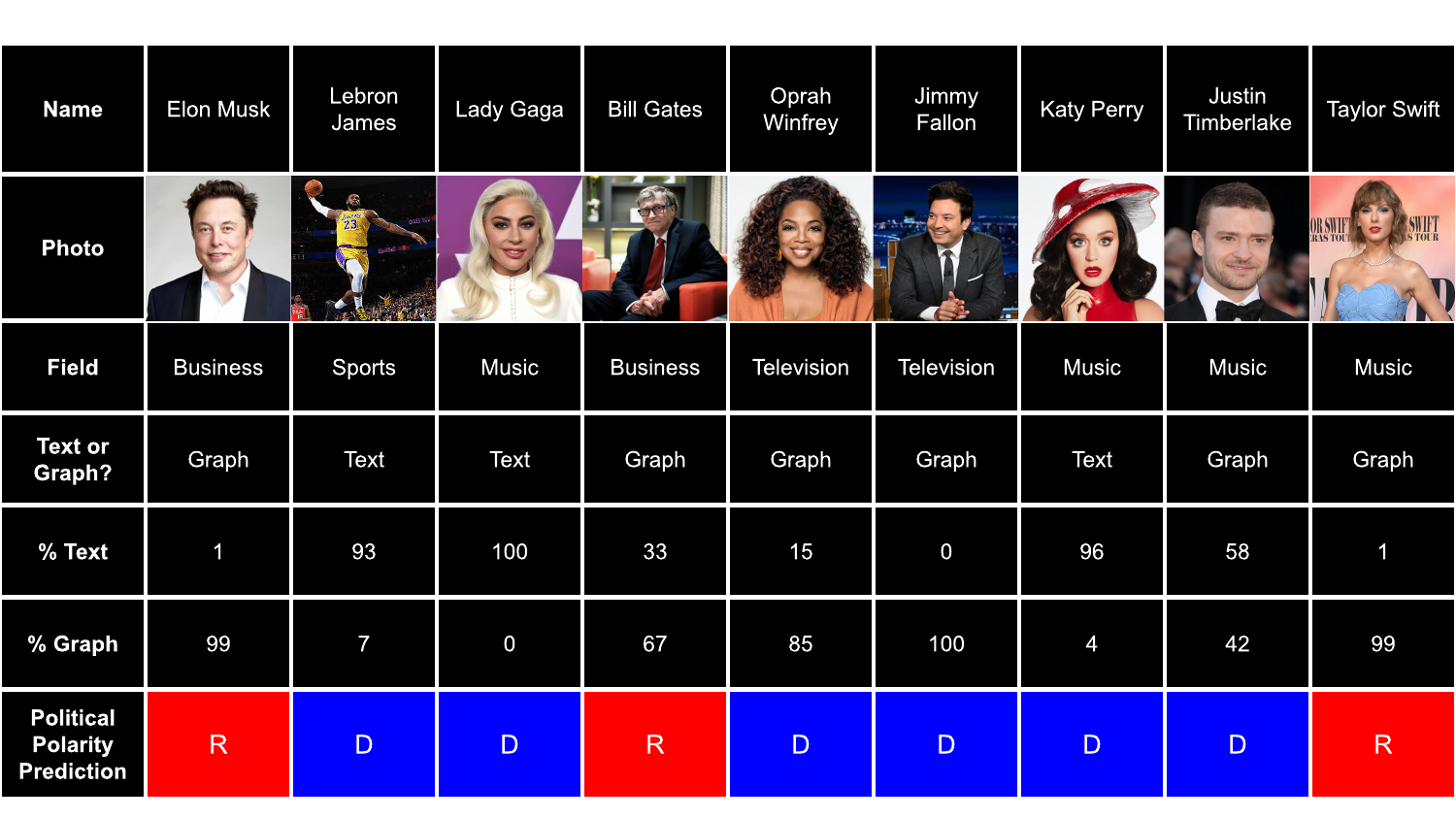} }}%
    \caption{A subset of the top 50 followed celebrities, note that the political polarity prediction is coming from our model and may not reflect their actual stances.}%
    \label{fig:celabrity}%
\end{figure*}

To conclude, we are able to obtain customized user behavior patterns through our multi-modality fusion. Those patterns could provide insights on which modality we should focus on more for different type of users, for downstream tasks such as personalized recommendations, social science analysis, or malicious user detection.
\\
\\
\textbf{Selected Celebrities from TIMME Dataset} Since we are not allowed to disclose regular Twitter users' information, we instead selected 9 celebrities among the top 50 most followed Twitter accounts from \footnote{\url{https://socialblade.com/twitter/top/100}}, whose Twitter accounts appear in the TIMME dataset, as a case study to show our frameworks' capability to give personalized explanations. We run the political polarity prediction task on those people and obtained some predictions. We also record the percentage of text information and percentage of graph structure information that contributes to their political polarity prediction (See Figure \ref{fig:celabrity}).

\begin{itemize}
    \item Elon Musk: Before 2020 (dataset cut-off), Elon Musk's political views in his tweet text content are often complex. He has claimed several times not to take the viewpoints in his tweets too seriously \footnote{\url{https://twitter.com/elonmusk/status/1007780580396683267}}. This aligns with the low contribution weight of his texts on his political stand prediction. However, on the graph level, 66.67\% of the politicians Elon Musk liked have liked Trump at least once, which is significantly larger than the average number in the TIMME dataset ($23.67\%$). This could be a strong reason why our graph structure weight is so high and why we predict Elon Musk to be Republican-leaning. Our prediction is proved correct when in 2022 (which is beyond our dataset cut-off time, 2020 \cite{xiao2023detecting}), Elon Musk claimed that he would vote for Republicans in his tweet \footnote{\url{https://twitter.com/elonmusk/status/1526997132858822658}}. This is a strong indicator that our framework is using correct information.
    \item LeBron James: In his tweets, LeBron James frequently shows his love and respect to Democratic President Obama \footnote{\url{https://twitter.com/KingJames/status/1290774046964101123}, \url{https://twitter.com/KingJames/status/1531837452591042561}}. Our prediction for him to be Democrat-leaning with a strong text contribution aligns with this observation. 
    \item Lady Gaga: Similarly to James, Lady Gaga also expresses explicitly in her tweets about her support of Democratic candidates \footnote{\url{https://twitter.com/ladygaga/status/1325120729130528768}}. Our graph weight becomes 0 for her case, meaning that using the text alone we could make sure she is Democrat-learning.
    \item Bill Gates: He usually avoids making explicit statements about whether he supports Democrats or Republicans in his tweets. Although our model predicts him as the Republican, the probability edge is very marginal (11\%).
    \item Oprah Winfrey: During the 2016 presidential campaign, she retweeted and mentioned her support for Democratic candidate Hillary Clinton frequently \footnote{\url{https://twitter.com/Oprah/status/780588770726993920}}, making the graph structure information a strong indicator of her Democratic stance.
    \item Jimmy Fallon: Jimmy Fallon has managed to maintain a sense of political neutrality in his tweets. His text contribution to the final prediction is 0. Even though the Twitter graph structure information indicates that he is Democrat-leaning, we still do not know for sure in real life whether he is a Democrat or Republican.
    \item Katy Perry: Just like Oprah Winfrey, Katy Perry also interacted with and supported Hillary Clinton during the 2016 election, a reason why we predict her as Democrat-leaning from the graph structure. Although she supports some republican politicians in 2022 \footnote{\url{https://twitter.com/katyperry/status/1533246681910628352}}, that is beyond the dataset cutoff.
    \item Justin Timberlake: Justin Timberlake has frequent positive interactions with President Obama \footnote{\url{https://twitter.com/jtimberlake/status/1025867320407846912}} and firmly supports Hillary Clinton in his tweets \footnote{\url{https://twitter.com/jtimberlake/status/768191007036891136}}, both suggesting that he is Democrat-learning. Our model assigns a similar weight to text and graph structure, suggesting that both of them are effective in making that prediction.
    \item Taylor Swift: In the case of Taylor Swift, the model fails to give the correct prediction. Her tweets show that she voted for Biden during 2020 \footnote{\url{https://twitter.com/taylorswift13/status/1266392274549776387}}, but the prediction is Republican. One reason is that at the graph structure level, the majority of Taylor Swift's followers are classified as Republican (67.09 \%) in the dataset, which could mislead the graph encoder.
\end{itemize} 

Overall, we find out that for those celebrities, graph structure information is usually more useful when making political polarity predictions. That aligns with the quantitative results in table \ref{tab:table12}. As we see, different celebrities could have very different behavior patterns, and those patterns could be correctly captured and explained by our contribution weight. That confirms the effectiveness of our framework.

\section{Conclusion}

In this paper, we investigate some potential limitations of existing fusion methods for text information and graph structure information in user representation learning from social networks. We then propose a contribution aware multimodal social-media user-embedding with a learnable attention module. Our framework can automatically determine the reliability of text and graph-structure information when learning user-embeddings. It filters out unreliable modalities for specific users across various downstream tasks. Since our framework is not bound to any specific model, it has great potential to be adapted to any graph-structure-embedding component and text-embedding component, if affordable. More importantly, our models can give a score on the reliability of different information modalities for each user. That gives our framework great capability for personalized downstream analysis and recommendation. Our work can bring research attention to identifying and removing misleading information modality due to differences in social network user
behavior, and paves the way for more explainable, reliable, and effective social media user representation learning.

Some possible future extensions include adding more modalities other than text and graphs (e.g., image and video data from user's posts). Also, we consider the user identities to be static throughout our analysis, which might not be the case in many scenarios. We can bring time as a factor to produce a multi-modality dynamic social media user embedding. For example, it is possible to observe that a user's text content is more trustworthy in the first few months, and then that user's interactive graph structure information or interaction becomes more reliable in longer-terms.

\bibliography{main}

\end{document}